# Strategies and tolerances of spin transfer torque switching

Dmitri E. Nikonov, George I. Bourianoff

Intel Corp., Components Research, Santa Clara, California, 95052

Graham Rowlands, Ilya N. Krivorotov

Department of Physics and Astronomy, University of California–Irvine, Irvine, California 92697, USA

dmitri.e.nikonov@intel.com

## Abstract

Schemes of switching nanomagnetic memories via the effect of spin torque with various polarizations of injected electrons are studied. Simulations based on macrospin and micromagnetic theories are performed and compared. We demonstrate that switching with perpendicularly polarized current by short pulses and free precession requires smaller time and energy than spin torque switching with collinear in plane spin polarization; it is also found to be superior to other kinds of memories. We study the tolerances of switching to the magnitude of current and pulse duration. An increased Gilbert damping is found to improve tolerances of perpendicular switching without increasing the threshold current, unlike in plane switching.

## *1. Introduction*

Research in spintronics resulted in huge technological impact via development of extremely high capacity hard drives and magnetic RAM[1]. The basic paradigm of such devices is a stack of two ferromagnetic (FM) layers separated by a non-magnetic layer, see Fig. 1a. Giant magnetoresistance[2,3] (GMR) – dependence of resistance of such a stack on the relative direction of magnetization in the FM layers – provides the crucial path to interfacing magnetic and electronic states in the device. If the non-magnetic layer is a dielectric, this effect is called tunneling magnetoresistance (TMR)[4,5,6,7].

A better way of switching magnetization in such devices was first theoretically predicted[8,9] followed by the demonstration[10,11,12,13,14] of spin transfer torque effect. This effect consists of precession of magnetization of one of the FM layers as current flows across the stack. As this happens, the angular momentum of spin-polarized current originating in one FM layer is transferred to the magnetization of another FM layer, see Fig 1a. Since this original work, a tremendous amount of research has been conducted in the field (see review[15]). Good values of the performance metrics have been achieved: the memory cell size has been reduced to a few square microns, the switching time to a few nanoseconds, and the switching current to a few milliamps. Spin transfer torque random access memory (STTRAM) has been prototyped[16] and is close to commercialization.

Still to be competitive with the incumbent memory technologies, such as SRAM, DRAM, and flash, STTRAM has to surpass them in the majority of a set of metrics (size, speed, energy). The fact that STTRAM is non-volatile is an important advantage, but would not alone ensure commercial success. For this reason it is necessary to devise ways to decrease STTRAM's threshold switching current and time, and thus the overall switching energy. One of the possible pathways to this end is conducting switching by short pulses rather than quasi-steady currents.

A previous theoretical treatment[17] of pulsed currents envisioned two pulses of opposite polarity without a gap between them. Experimental demonstration of switching with pulses was performed with single pulses [18,19] or double pulses [20] of constant polarity.

In the present work we attempt to give a comprehensive view of the pulsed switching of nanomagnetic memories. We consider a variety of cases of spin polarization of the current injected from the fixed FM layer to the free FM layer. In considering various dynamical switching strategies, we first define the energy landscape of the system and represent a specific switching strategy as a specific path across the topological surface which defines the energy landscape   The analysis contained here identifies the optimal switching strategy to be along the path of steepest accent (unlike the traditional, in-plane collinear polarization switching which proceeds along the path of minimum accent). We perform simulations both in the macrospin and micromagnetic approximations and compare them side by side in order to make conclusions on applicability of these methods in each of the cases. In these simulations we separate the contribution of spin transfer and field-like torques [21,22,23] to draw a conclusion on how they affect switching in each of the cases. Finally we perform multiple simulations over a range of switching parameters in order to predict the tolerances of the memories relative to the variations in both the magnitude of current and the pulse duration. Such variations of current and pulse duration are bound to exist in electronic circuits due to fabrication variability and the temperature drift. Counter-intuitively, we find that Gilbert damping is beneficial for the tolerance of switching and increases neither the threshold current nor the switching duration.

The paper is organized as follows. Section 2 contains the description of the mathematical models of the macrospin approximation and micromagnetic simulator OOMMF used in this paper. Section 3 analyzes the energy dependence on the direction of magnetization and various strategies for switching based on it. Time dependence of magnetization in various strategies of switching is also exemplified. Results of multiple simulation runs, presented in Section 4, establish tolerances of switching relative to current magnitude and pulse duration. In Section 5 we compare the results of macrospin and micromagnetic simulations side by side. The conclusions of this work are summarized in Section 6.

## *2. Mathematical models*

The mathematical model of macrospin dynamics is based on the Landau-Lifshitz-Gilbert (LLG) equation (see the review[24]). In addition it assumes that spatial variation of magnetization can be neglected, and the whole magnetic moment of the nanomagnet ("macrospin") can be represented by a single average vector of magnetization $\mathbf{M}$, or dimensionless magnetization $\mathbf{m} = \mathbf{M}/M_s$, where the saturation magnetization of the material is $M_s$. Thus the LLG equation, containing the spin torque terms $\boldsymbol{\Gamma}$, is

$$\frac{d\mathbf{m}}{dt} = -\gamma\mu_0 \left[\mathbf{m} \times \mathbf{H}_{eff}\right] + \alpha\left[\mathbf{m} \times \frac{d\mathbf{m}}{dt}\right] + \mathbf{\Gamma}, \quad (1)$$

where the gyromagnetic constant is $\gamma = \dfrac{g\mu_B}{\hbar}$, the Bohr magneton is $\mu_B$, and the Lande factor is $g$, the permeability of vacuum is $\mu_0$, and the Gilbert damping factor is $\alpha$. The effective magnetic field originates from all contributions to the energy $E$ per unit volume of the nanomagnet:

$$\mathbf{H}_{eff} = -\frac{1}{\mu_0}\frac{\delta E}{\delta \mathbf{M}} \quad (2)$$

The energy of the nanomagnet includes the demagnetization term – coming from the interaction of magnetic dipoles between themselves, the material anisotropy, and the Zeeman energy due to an external magnetic field. In this article we disregard the latter two contributions for the free FM layer, and focus on the former. The demagnetization term (synonymous with the shape anisotropy of the nanomagnet) is

$$E = \frac{\mu_0}{2}\mathbf{M}\mathbf{N}\mathbf{M}, \quad (3)$$

The demagnetization tensor is diagonal and $N_{xx} + N_{yy} + N_{zz} = 1$, if the coordinate axes coincide with the principal axes of the nanomagnet:

$$\mathbf{N} = \begin{pmatrix} N_{xx} & 0 & 0 \\ 0 & N_{yy} & 0 \\ 0 & 0 & N_{zz} \end{pmatrix}. \tag{4}$$

For the shape of typical nanomagnets, elliptical cylinder, the demagnetization tensor is calculated according to Ref. [25]. A heuristic rule for demagnetization energy is that it is lowest when magnetization points along the longest axis of a nanomagnet and highest when it points along the shortest axis. For example, in the case considered here of the nanomagnet with dimensions of 120nm*60nm*3nm, the demagnetization tensor elements are $N_{xx}=0.0279$, $N_{yy}=0.0731$, $N_{zz}=0.8990$.

The spin torque contribution is described by the spin transfer (Slonczewski) and field like terms (in the brackets of the following equation)

$$\mathbf{\Gamma} = \frac{\gamma \hbar J}{M_s e t} \left( \varepsilon \left[ \mathbf{m} \times [\mathbf{p} \times \mathbf{m}] \right] + \varepsilon' [\mathbf{p} \times \mathbf{m}] \right), \tag{5}$$

where $t$ is the thickness of the nanomagnet, e is the absolute value of the electron charge, $J$ is the density of current perpendicular to the plane of the nanomagnet, $\mathbf{p}$ is the unit vector of polarization of electrons, and $P$ is the degree of polarization of these electrons. In this paper we disregard the angular dependence in the prefactor for simplicity, so that

$$\varepsilon = \frac{P}{2} \tag{6}$$

The mathematical model of micromagnetics is realized in a widely used simulator OOMMF [26]. It is based on the same LLG equation (1). The main difference from the macrospin model is that now magnetization **M** is considered a function of spatial coordinates (discretized on a grid) and its spatial variation plays an important role. The exchange interaction of between spins is included as the energy density

$$E = A\left(\left|\nabla m_x\right|^2 + \left|\nabla m_y\right|^2 + \left|\nabla m_z\right|^2\right) \quad (7)$$

and the demagnetization energy is calculated by explicit summation of dipole-dipole interactions between the parts of the nanomagnet, see Ref. 26. In this paper we will use the following set of typical parameters and hope that the reader agrees that the conclusions of the paper do not depend on this particular choice of numerical values: saturation magnetization $M_s = 1MA/m$, Lande factor $g = 2$, spin polarization $P = 0.8$, and the exchange constant $A = 2 \cdot 10^{-11} J/m$. In the cases when we include the field-like torque, we set its constant $\varepsilon' = 0.3\varepsilon$ to be in approximate agreement with the results of Refs. 21 and 22.

## *3. Energy profile and strategies for switching*

We are applying the above mathematical model to treat various cases of spin transfer torque switching, Fig. 1. In these schemes of nanopillars, the upper blue layer designates a free nanomagnet. We introduce the coordinate axes as follows: x along the long axis of the nanomagnet, in plane of the chip, y perpendicular to x in plane of the chip, and z perpendicular to the plane of the chip. We also introduce the angles to specify the magnetization direction: $\theta$, the angle from the x-axis, and $\phi$, the angle of the projection on the yz-plane from the y axis. The free nanomagnet has two stable (lowest energy) states when magnetization points along +x or –x directions, i.e. $\theta = 0, \pi$. The goal of memory engineering is to switch magnetization between these two states. The bottom blue layer designates the fixed nanomagnet. Even though the spin torque acts on this layer as well, one keeps its magnetization from switching by coupling it to an adjacent ("pinning") antiferromagnetic layer (not shown in the picture). The magnetization of the fixed nanomagnet can be set in various directions by fabricating it with the right shape and magnetocrystalline anisotropy. One case is when the fixed magnetization is approximately along the x-axis, Fig. 1a. If both magnetizations were exactly along the x-axis, the spin torque acting on the free layer would be zero, and switching would not take place. In fact, thermal fluctuations cause the angles of both free and fixed nanomagnets to have random values around $\theta = 0$. Therefore in the macrospin simulations, we formally set the initial angle of the free nanomagnet to $\theta = 0.1$ radians, and set the angle of the fixed layer to $\theta = 0$. Another case, Fig 1b, is that of the fixed magnetization in plane of the chip, with various $\theta$ and $\phi = 0$. Finally, the case in Fig. 1c is that of perpendicular magnetization of the fixed nanomagnet $\phi = \pm \pi/2$. We will see further on that the directions of magnetization of the fixed layer, which we consider identical to directions of spin polarization of the electrons injected into the free layer, correspond to different strategies of switching.

In order to gain an intuitive understanding of the process of switching, one needs to visualize the "energy landscape" – the pattern of demagnetization energy in the phase space of magnetization angles, according to Eq. (3). The map of this sphere of angles on the plane is shown in Fig. 2. For a different shape of the nanomagnet, or in the presence of material anisotropy and external field, this dependence will be quantitatively different, but the same qualitative approach applies. The salient features of the energy landscape are: a) the stretched ellipse-shaped "basins" close to the poles – the stable equilibrium states; b) the two "valleys" stretching from one pole to the other – the states with in-plane magnetization; c) they have "mountain passes", or saddle points at $\theta = \pi/2$ and $\phi = 0, \pi$ d) two peaks corresponding to magnetization perpendicular to the plane, at $\theta = \pi/2$ and $\phi = \pm\pi/2$. The iso-energy lines are shown in the contour plot, Fig. 2. In the absence of damping and spin torque, they would coincide with closed orbits of magnetization. There are orbits of low energy precessing around one of the poles, and orbits of high energy oscillating between the two poles. In the presence of damping, the nanomagnet will evolve to one of the basins and eventually to the pole inside it. Spin torque can cause the nanomagnet to gain or lose energy, under some conditions moving between the basins, i.e. switching.

At this point, let us agree on the definitions for switching time. We will consider switching under the action of rectangular pulse currents of magnitude $I$ and duration $\tau_{pu}$ (marked in the following plots). These values are relevant for the switching charge $I\tau_{pu}$ and energy $E_{sw} = UI\tau_{pu}$ dissipated in switching under the voltage bias $U$. Therefore the pulse duration $\tau_{pu}$ is the time important in the technological sense for optimizing the energy per writing a bit. From simulations we obtain the switching time, which characterizes how fast the nanomagnet responds to the current pulse. We customarily define the switching time $\tau_{sw}$ as the time over which the magnetization is switched from 10% to 90% of its limiting values. In our particular case it is the interval between the first time the projection of magnetization $m_x$ goes below 0.8 till the last time it is over - 0.8. The 10-90% time gives the lower bound of switching time. Its importance is to characterize the switching time pertinent to the strategy, rather than influence of initial conditions. The reason that we use it instead of 0-100% time is that, in the collinear in-plane case, the latter strongly depends on the choice of the initial angle of magnetization (see discussion below). The total write time needs to be longer than the largest of the two time measures.

In the following, we will plot the switching times $\tau_{sw}$ resulting from the simulation of magnetization dynamics over certain time intervals, typically 1, 2.5, 3, or 4ns. When the switching time reaches this constant value, it really means that switching does not occur over the simulation time, and, with high probability, even after any duration of evolution. Thus these limiting constant values on the plot are just tokens for "no switching occurring".

Collinear polarization spin torque switching is done in a configuration of Fig 1a. Since the injected spin polarization is in plane close to $\theta = 0$, the resulting spin transfer torque pushes the magnetization to rotate in plane, along the slope of slowest accent in energy, which might seem at first glance like the optimal strategy of switching. An example of switching dynamics for this case is shown in Fig. 3. From the time evolution plot we see that magnetization performs an oscillatory motion close to in plane position with slowly increasing amplitude. Torque increases with the angle from the x-axis, and this reinforces the growth of this angle. At some point the projections on the x-axis abruptly switches to a negative value and then the amplitude of the oscillations is damped[27,28]. From the trajectory in phase space of magnetization direction, we see that the switching happens by moving along one of the valleys and crossing over a saddle point. The duration of the pulse is sufficient if it is longer than the time necessary to go to the other side of this saddle point.

Another strategy is the in the configuration of Fig. 1b. It is similar to the previous strategy, with a few modifications. In case of the spin polarization 90 degree in plane, i.e. $\theta = \pi/2$, see Fig. 4, the torque is maximal in the initial instant when the magnetization is at $\theta = 0$. At very large current values, when the magnetization reaches the saddle point, the torque turns to zero and the nanomagnet dwells in an unstable equilibrium until the end of the current pulse. At this point it falls towards one of the equilibriums; the choice of which is governed by the randomness of its position at that moment. Overall, this looks like an unreliable method of switching. Also it requires a much higher current than collinear polarization switching and the switching time is longer[29]. The situation is drastically improved for the case of injected spin polarization at a different angle, e.g. 135 degree in plane, $\theta = 3\pi/4$, shown in Fig. 5. The path of switching still goes along the energy valley and over the saddle point. But in that case, torque is not zero at the initial instant, and it does not turn to zero at the position of the energy saddle point. Switching time proves to be shorter. But one important similarity is that the pulse time needs to be relatively long in order to cross over the saddle point. The strategy of switching with perpendicular spin polarization, as shown in Fig 1c, turns out to be very different from collinear polarization switching. It stems from the fact that the spin transfer torque acts in the direction perpendicular to the sample plane as well. In a counter-intuitive manner, it pushes the nanomagnet along the path of steepest accent. The way to take advantage of this situation is to use a very short pulse, which will supply sufficient energy to the nanomagnet. The advantage of a short pulse is that it requires smaller switching energy supplied by the current. After the initial short current pulse, the nanomagnet precesses due to the torque from the shape anisotropy at zero spin torque, see Fig 6 for an example of such evolution. Under the condition of a small Gilbert damping, its trajectory will be close to the iso-energy line. A necessary condition for switching is that the nanomagnet has sufficient energy to be on the trajectory that crosses to the other basin, even with the account of loss to damping. The sufficient condition of successful switching is that by the time magnetization reaches the other basin, it loses enough energy so that it cannot cross back to the original basin. If this condition is satisfied, the nanomagnet slowly loses energy to damping and approaches the equilibrium. Due to the small value of damping the switching time turns out to be long.

A variation on this strategy is to apply another pulse of current after the nanomagnet crosses to the other basin, Fig. 7. This pulse can have the same duration $\tau_{pu}$ and start after a delay time, $\tau_g$, after the trailing edge of the first pulse but must have the opposite polarity of the current. Such a two-pulse sequence with the total time $\tau_t = 2\tau_{pu} + \tau_g$, will efficiently decrease the energy and avoid the process of slow energy damping. As a result the switching time $\tau_{sw}$ becomes very short, ~0.2ns, comparable to $\tau_{pu}$. The downside of this strategy is that two pulses of course require twice the energy of one pulse with the same current magnitude and duration. Also it requires a more complicated circuit to time the pulses of opposite polarity. We note that the difference of the strategy considered here from the one of Kent et al.[17] is that in their approach the two pulses of the opposite polarity did not have a gap between them, so the stage of free precession was absent.

We make the following approximate estimate of the pulse parameters for the pulsed switching. The energy that the nanomagnet gains in the first pulse must be larger than the energy necessary to cross over the saddle point.

$$m_z^2 N_{zz} = N_{yy} - N_{xx} \qquad (8)$$

On the other hand, the out of plane projection at the end of the pulse is obtained from Eq.

$$m_z = \frac{g\mu_B I \tau_{pu} P}{2 M_s eV}. \qquad (9)$$

For the parameters used in this paper, it amounts to the switching charge of $I\tau_{pu} = 81 fC$ per bit. This is much smaller than the best achieved values for the collinear polarization switching, ~5pC per bit. Moreover, if we compare the switching time and energy projected here with incumbent types of memory, see Table 3 in Ref. [30], we see that STTRAM with pulsed switching is superior to all other types of memory. From prototypes of STTRAM [16] we also know that its density can be comparable to that of DRAM. Therefore the proposed improvement gives it the crucial performance boost to potentially be the universal memory and to replace all other kinds.

### *4. Tolerances of switching and influence of field-like torque*

Spin transfer torque memories operate in electronic circuits. The circuits naturally have variability originating from fabrication imperfections. Also the state of the circuit is subject to electronic noise and temperature drift. Obviously it is not possible to guarantee precise values of switching current and time for elements of memories. Therefore it is especially important to study the tolerances of memory operation relative to external parameters. We believe such studies have not been conducted up to now. Here we run multiple simulations over a wide set of parameters to draw some conclusions about these tolerances.

At the same time we study the effect of the field-like torque (FLT) contribution. Experiments on separate measurements of the spin transfer (Slonczewski) and field like torques have been conducted [21,22]. But the implications of these contributions to current induced magnetization switching have not been sufficiently clarified. The experiments show that FLT increases with increasing applied voltage. To account for this we consider two cases – no field-like torque $\varepsilon' = 0$ and large field-like torque $\varepsilon' = 0.3\varepsilon$.

The contour plots of switching time vs. current and pulse duration for collinear polarization switching are shown in Fig. 8. It is common to think of spin torque switching as having a threshold (or critical) current $I_c$. However from this plot we see that, for sufficiently short pulse duration, the switching current $I - I_c \sim 1/\tau_{pu}$ is much larger than the critical current. Therefore it is the threshold charge, $I\tau_{pu} \approx 1.2\,pC$, that determines whether the memory state is switched. Above this threshold, switching can be quite fast, ~0.2ns, but the total write time is limited by the pulse duration instead. One can notice that the shapes of the switching time dependence with and without FLT are remarkably similar, but they appear to be shifted. For that reason, one needs to be cautious of the fact that for a specific values of current and pulse duration, the dynamics may be different with and without FLT. The reason for similarity is that for this case FLT plays a role of an effective magnetic field in the z-direction, in addition to a large effective field from demagnetization. There are curious geometrical features of the switching threshold border. Even though they are persistent in simulation, we believe that they are artifacts of the choice of initial magnetization and of oscillations ("ringing") of magnetization after switching. In reality, thermal fluctuation will vary the initial magnetization, and the features on the plot will be washed away for the thermally averaged switching time. Overall, this strategy of switching gives an excellent tolerance when the switching current and pulse duration are set high enough above the threshold.

The switching time diagrams for the 90 degree in plane polarization are shown in Fig. 9. In this case, the threshold current is actually a good criterion of switching; and this threshold turns out to be very high, ~10mA. Without FLT, even above threshold there are tightly interlaced regions of successful and unsuccessful switching. This attests to the unstable nature of such switching. It cannot be used for a practical device. The situation is different with FLT. There are large regions of small switching time, and therefore good tolerance to parameters, above the threshold. The reason for this stabilization is that even though the Slonczewski torque vanishes at $\theta = \pi/2$, FLT is still finite and it succeeds in pushing the nanomagnet over the energy saddle point. However at excessively high current we encounter the regions of unsuccessful switching that memory designers need to avoid.

The switching diagrams for 135degree in plane polarization are shown in Fig 10. The threshold behavior is in between the collinear polarization and 90degree in plane cases. The threshold current is not constant, and it is ~2-4mA, which is lower than that for 90 degrees. But it is not inversely proportional to the pulse duration either. The threshold charges are in the range $I\tau_{pu} \approx 0.6 \div 1.6\, pC$. Overall it has the same excellent tolerance to current and pulse variation as the collinear polarization switching, but in the absence of FLT, regions of unsuccessful switching are observed at higher current.

The switching diagrams for out of plane spin polarization and one switching pulse are shown in Fig. 11. The areas of successful switching are seen as narrow strips across the plot. They are interlaced with stripes of unsuccessful switching. The reason for such behavior is the precession character of magnetization dynamics for this switching strategy. If too much energy is transferred from the current to the nanomagnet, it overshoots and returns to the basin around the initial magnetization state. The set of successful switching correspond to 0.5, 1.5, 2.5 etc. full turns of magnetization. The lowest of these stripes corresponds to the threshold of switching, $I\tau_{pu} \approx 100\, fC$. Though the threshold is only approximately given by the product of the current and the pulse duration. This is in a very good agreement with the analytical estimate (9).

This is the first case when we encounter the problem of tolerances in earnest. From the left plot in Fig. 11 for Gilbert damping of 0.01, we estimate the tolerances to be 1ps and 0.1mA. This is likely too tight for a realistic memory circuit. The right plot is calculated for Gilbert damping 0.03. We see that, contrary to what is known about collinear polarization switching, the threshold is almost unchanged at a higher Gilbert damping. At the same time, the tolerances are much relaxed, to 5ps and 0.7mA. The switching stripes became wider, and the next order of switching with 1.5 turns is moved to a higher switching charge. This seems to be the first occasion when increasing damping is beneficial for the device performance. We note that these simulations are done with inclusion of FLT. The results with zero FLT (not included in this paper) are almost indistinguishable from those. The reason for this is that the current pulses act when magnetization is close to the poles, and FLT has projections mostly in-plane of the nanomagnet, which contribute only negligibly to the precessional type of switching.

The switching diagrams for out of plane polarization and two pulses are shown in Fig. 12. The total pulse time is fixed at $\tau_t = 0.2 ns$, while the pulse time $\tau_{pu}$ is given on the horizontal axis of the plot. Their overall character is similar to those for a single pulse. The threshold condition is approximately the same. The first stripe is very narrow, with the tolerances 0.3ps and 30uA. Surprisingly the second stripe corresponding to 1.5 turns, has a much larger and acceptable value of tolerance of 4ps and 0.4mA. We do not have an intuitive explanation for this difference between the two switching cases. For smaller Gilbert damping 0.01 the switching time is quite short, ~0.2ps. This is due to the fact that the second pulse eliminates ringing of magnetization, as it was discussed in the previous section. With increase of Gilbert damping to 0.03, the tolerances in the first stripe improve to 2ps and 0.2mA. This is manifested as appearances of several satellite strips around the first one, meaning that at higher damping the condition of timing the gap between the pulses becomes less crucial for successful switching. Conversely, the switching time gets slower, ~0.5ps. This is because the magnetization oscillations are not eliminated as efficiently is the pulses are not finely timed. Like for the single pulse, the inclusion of FLT changes the results very insignificantly.

## 5. Comparison of macrospin and micromagnetic simulations

The simulation of micromagnetic dynamics is a more rigorous model and presumably gives a better approximation to reality that the macrospin approximation. However it is also much more computationally demanding. For this reason it is useful to compare the results of macrospin and micromagnetic models. We compare the switching time at various pulse durations but fixed values of current. Micromagnetic simulations start with an initial state obtained by relaxing its energy to a minimum. It has the general direction of $\theta = 0$, and a "leaf state" pattern, i.e. magnetization is closer to being parallel to the longer sides of the ellipse.

The comparison for the collinear polarization switching is shown in Fig. 13. Here we set the direction of the spin polarization of the fixed layer to be $\theta = 0.1$. This value is an estimate of the r.m.s deviation of the angle of magnetization due to thermal distribution of energy. We make this choice of the initial angle only in the case of collinear in-plane switching, because spin torque would vanish for $\theta = 0$. For other cases the spin torque does not vanish at the zero initial angle. The agreement is surprisingly good. Both models give approximately the same value of threshold charge. This may be because we are focusing on the evolution starting from angle $\theta = 0.1$. The evolution around $\theta = 0$ occurs under a much smaller torque and carries more uncertainty. Macrospin model exhibits more oscillations close to the threshold. Micromagnetic model predicts shorter switching time above threshold, probably due to more efficient damping of higher modes of magnetization.

The comparison in the case of 90degree in plane polarization, Fig. 14, shows essentially the same lower envelope of switching time. The micromagnetic model shows fewer areas of failed switching. We speculate that this is due to stronger effect of FLT on non-uniform magnetization.

The disagreement is more pronounced in the case of single pulse out of plane polarization, Fig. 15. The similarities are the same position on the time scale of successful switching stripes for 0.5 and 3.5 turns. The regions of unsuccessful switching for 1 and 3 turns are absent in the micromagnetic model, probably because the barrier for crossing into the other basin is increased for a non-uniform magnetization pattern. The switching times are generally predicted to be shorter in the micromagnetic model.

A similar situation is observed in out of plane polarization switching with two pulses, see Fig. 16. The positions of successful switching on the time scale are shifted. This is probably due to the fact that the nanomagnet energy is different with the account of exchange and dipole-dipole interaction, and thus the time of free precession of magnetization is different too. As before, micromagnetics predict shorter switching times, as well as giving better tolerance for 0.5 turn switching. In fact, micromagnetic simulations for one and two pulses look quite similar. This points to higher significance of damping than of the second pulse in bringing the magnetization to its final state.

This paper includes just a few examples of comparison. A more complete set of data is available [31]. All of this supports the conclusion that macrospin simulations give generally the same qualitative dependence of switching time on the current and pulse duration as OOMMF simulations. The former can be used to infer general trends. The latter should be saved for obtaining quantitatively precise results.

## *6. Conclusions*

We have compared various strategies of spin torque switching. The tolerances of switching, performance limits of out-of plane polarization devices, and the effect of field-like torque have been comprehensively studied for the first time. In summary, switching with short pulses of out of plane polarization is the preferred strategy. It has a much lower threshold charge than other strategies of switching, and suffers less from low tolerance to current magnitude and pulse duration. This problem of low tolerance can be resolved by increasing Gilbert damping in the nanomagnet. Field-like torque significantly changes the results for non-collinear in-plane switching and happens to produce a minor effect for other switching strategies. Implementation of this strategy would put STTRAM in a position of a technological leadership among memories. We find that macrospin gives good qualitative prediction of the dynamics, though micromagnetic models should be used to get better quantitative precision.

## *8. Acknowledgements*

G. R. and I. N. K. gratefully acknowledge the support of DARPA, NSF (grants DMR-0748810 and ECCS-0701458) and the Nanoelectronic Research Initiative through the Western Institute of Nanoelectronics. Computations supporting this paper were performed on the BDUC Compute Cluster donated by Broadcom, co-run by the UCI OIT and the Bren School of Information and Computer Sciences.

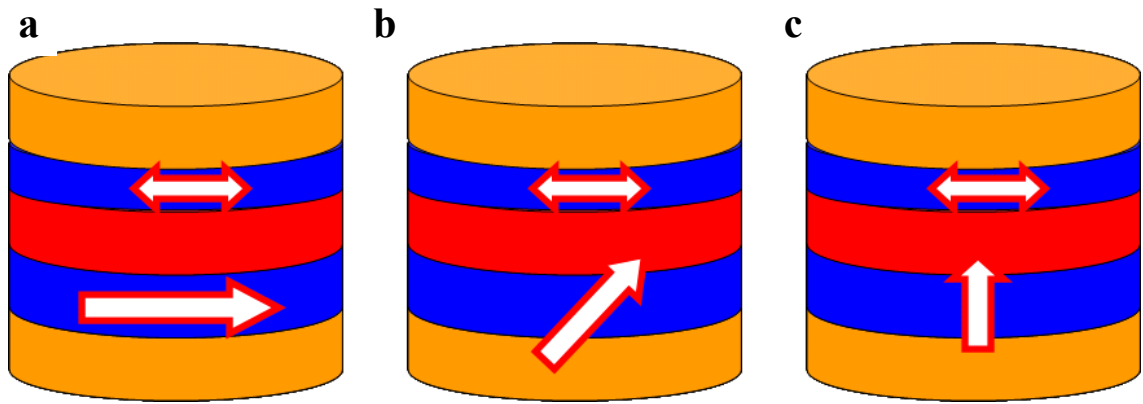

Figure 1. The geometry of the spin torque memory nanopillars. Red layer – tunneling barrier oxide, orange – non-magnetic electrodes, blue with double-sided arrow – free layer with in-plane magnetization (easy x-axis), blue with one sided arrow – fixed layer. Polarizations: a) in plane collinear with x-axis (left), b) in plane at an angle to x-axis (middle), c) perpendicular to the plane (right).

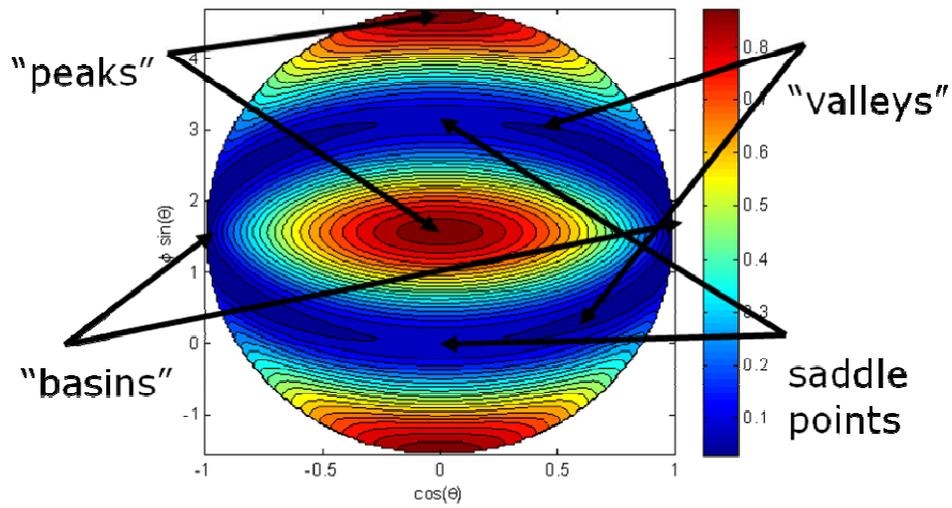

Figure 2. Map of the demagnetization energy of the nanomagnet (normalized) in the macrospin model. $\theta$ – angle of magnetization from the x-axis (easy axis), $\phi$ – angle of projection of magnetization within the yz-plane (hard plane).

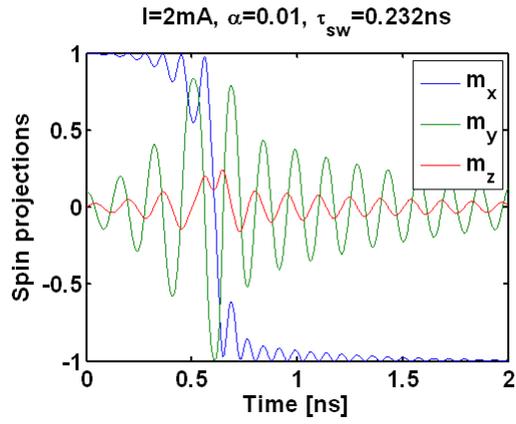 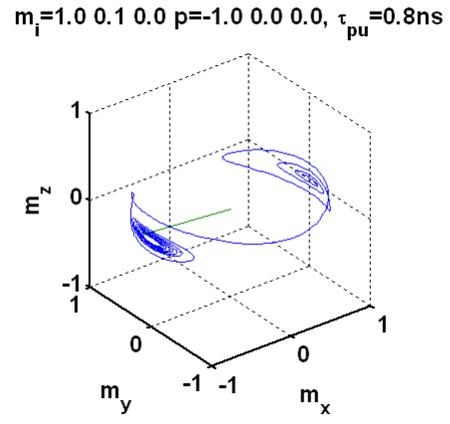

Figure 3. Magnetization projections vs. time and the trajectory of magnetization for collinear polarization switching (0 degree in plane polarization).

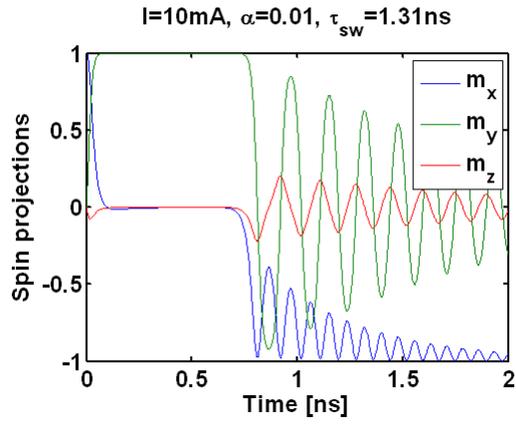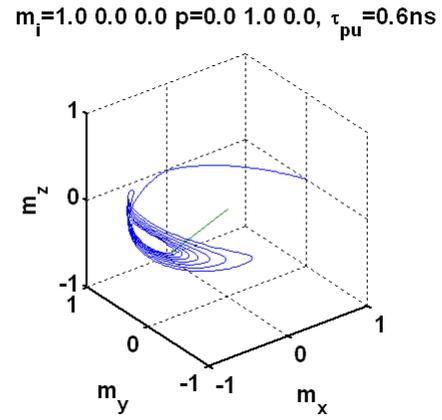

Figure 4. Magnetization projections vs. time and the trajectory of magnetization for 90 degree in plane polarization.

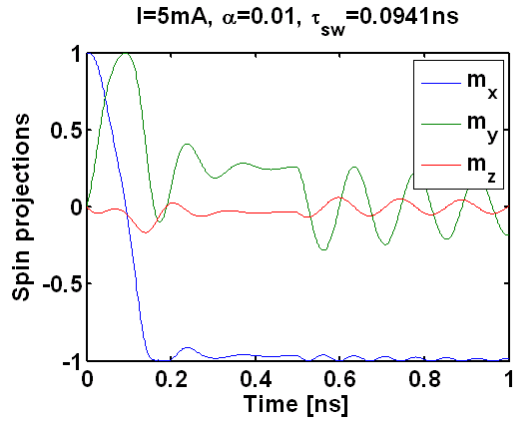 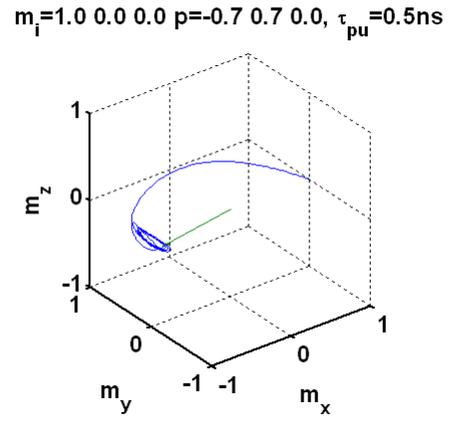

Figure 5. Magnetization projection vs. time and the trajectory of magnetization for 135 degree in plane polarization.

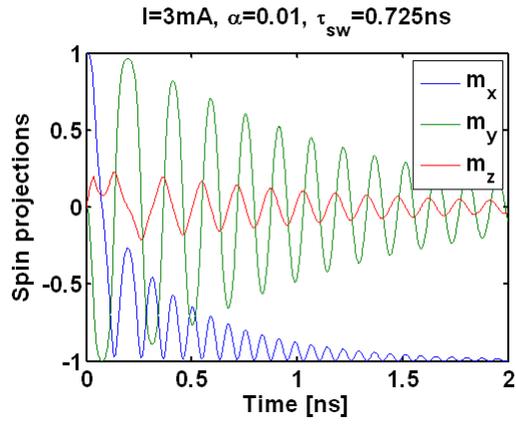 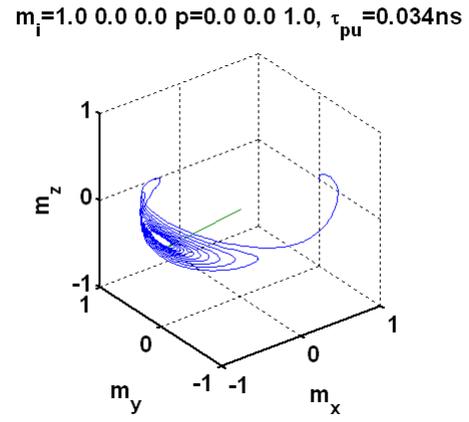

Figure 6. Magnetization projections vs. time and the trajectory of magnetization for perpendicular out of plane polarization, one pulse.

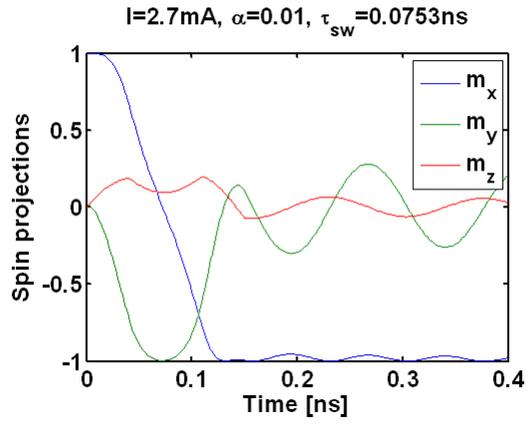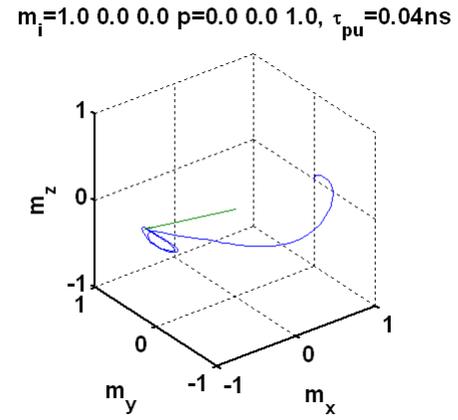

Figure 7. Magnetization projections vs. time and the trajectory of magnetization for perpendicular out of plane polarization, two pulses with a total time of 0.2ns.

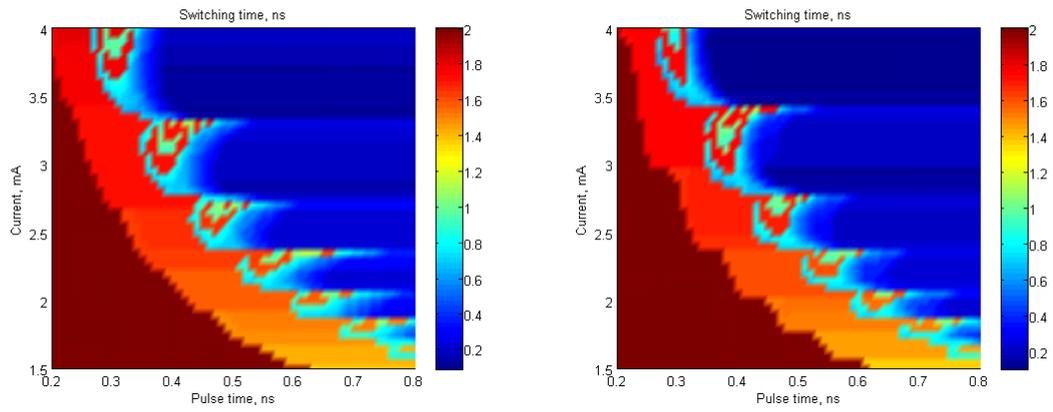

Figure 8. Contour maps of switching time, left without the field like torque and right with 0.3 factor of field like torque, for collinear polarization switching (0 degree in plane polarization). Simulation time 2ns.

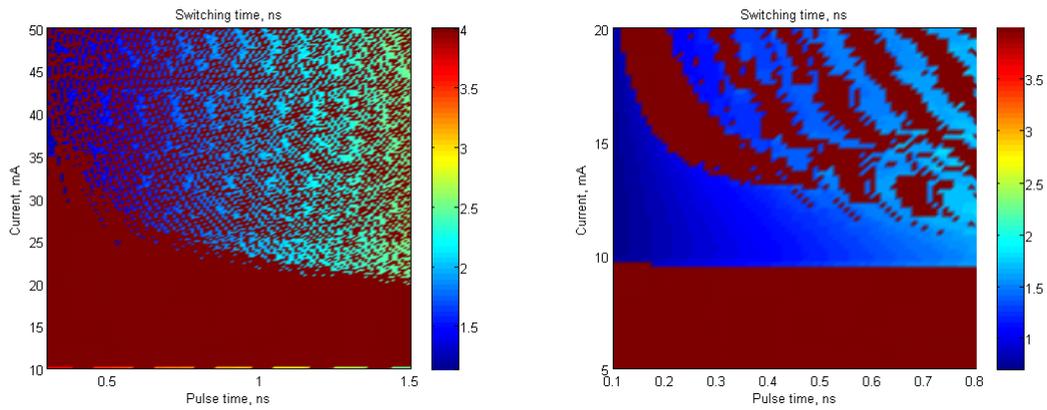

Figure 9. Contour maps of switching time, left without the field like torque and right with 0.3 factor of field like torque, for 90 degree in plane polarization. Simulation time 4ns.

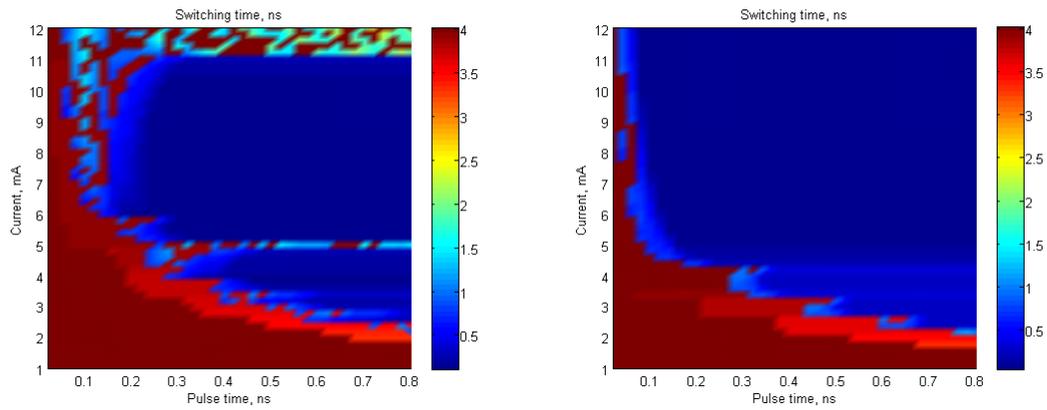

Figure 10. Contour maps of switching time, left without the field like torque and right with 0.3 factor of field like torque, for 135 degree in plane polarization. Simulation time 4ns.

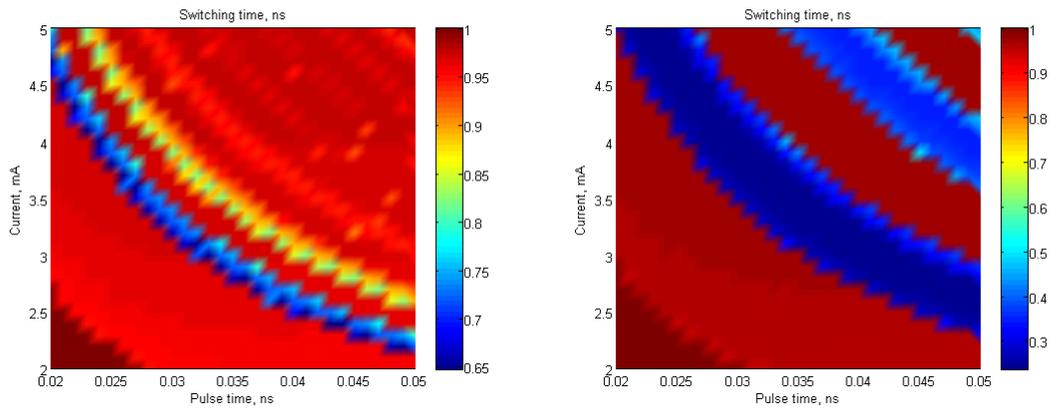

Figure 11. Contour maps of switching time, left $\alpha=0.01$ and right $\alpha=0.03$, with 0.3 factor of field like torque, for perpendicular out of plane polarization, one pulse. Simulation time 1ns.

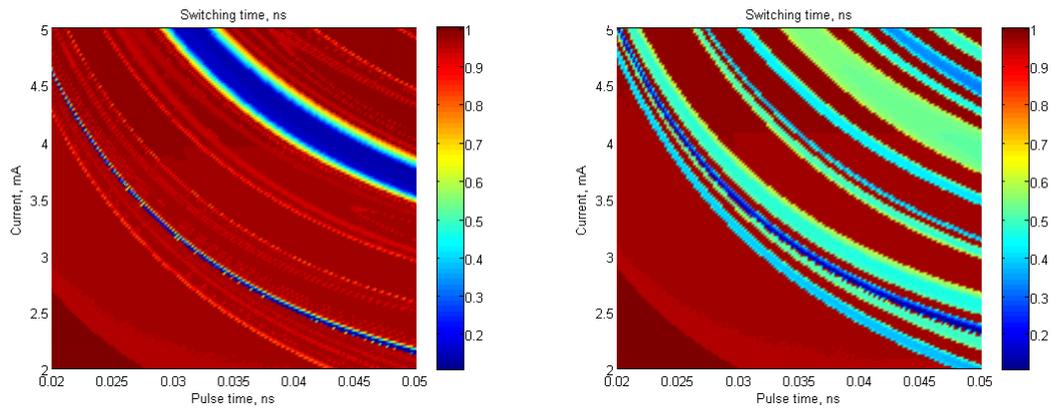

Figure 12. Contour maps of switching time, left $\alpha=0.01$ and right $\alpha=0.03$, with 0.3 factor of field like torque, for perpendicular out of plane polarization, two pulses with a total time of 0.2ns. Simulation time 1 ns.

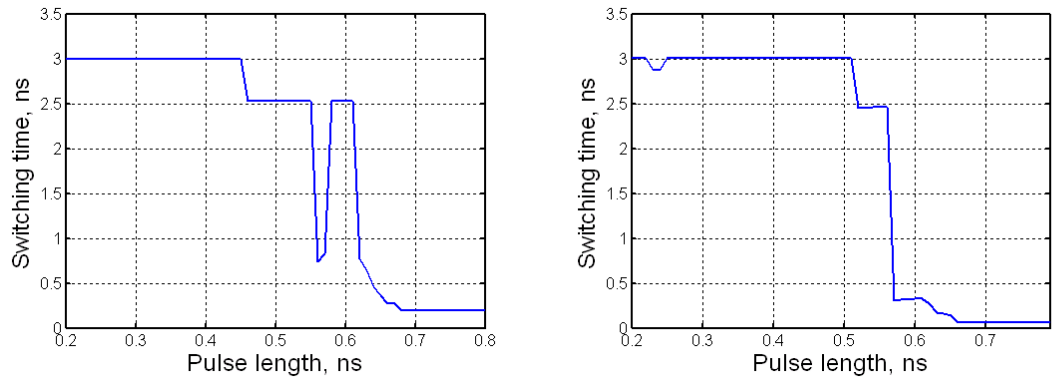

Figure 13. Comparison of the switching time vs. pulse length simulated by a macrospin model (left) and OOMMF (right) for collinear polarization switching at I=2.5mA, field-like torque 0.3, and polarization angle in plane of 10 degrees. Simulation time 3ns.

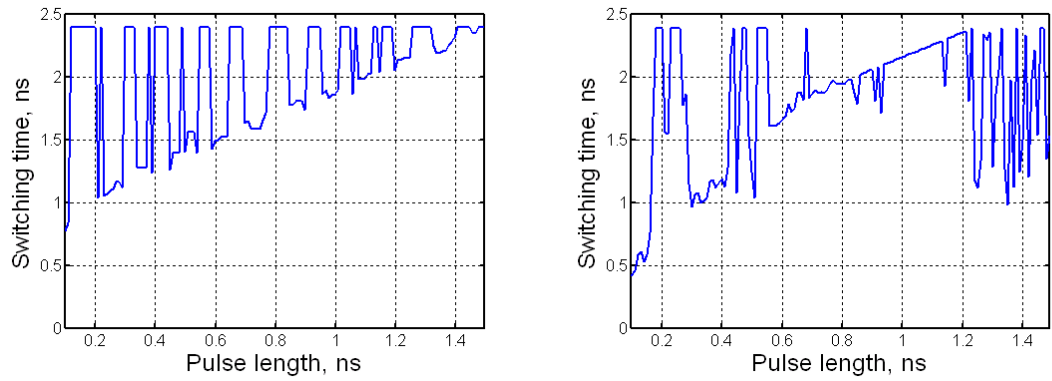

Figure 14. Same as Fig. 13 at I=20mA, field-like torque 0.3, and polarization angle in plane of 90 degrees. Simulation time 2.4ns.

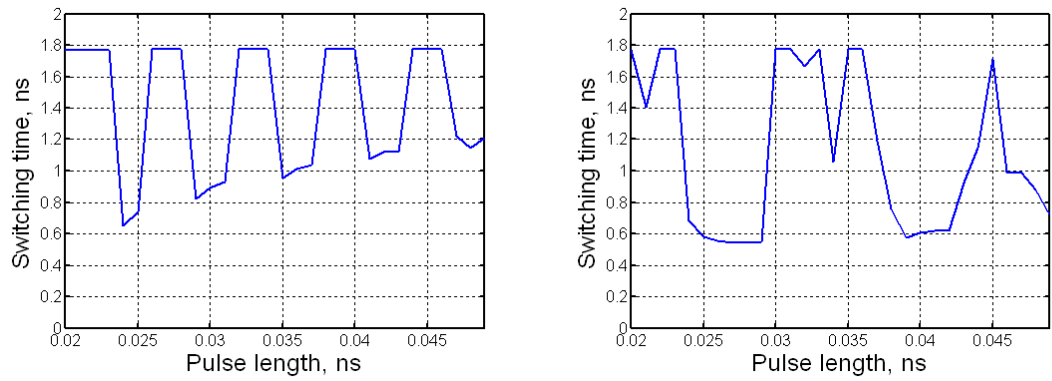

Figure 15. Same as Fig. 13, single pulse at I=4mA, field-like torque 0.3, $\alpha$=0.01, and polarization angle perpendicular to plane of 90 degrees. Simulation time 1.8ns,

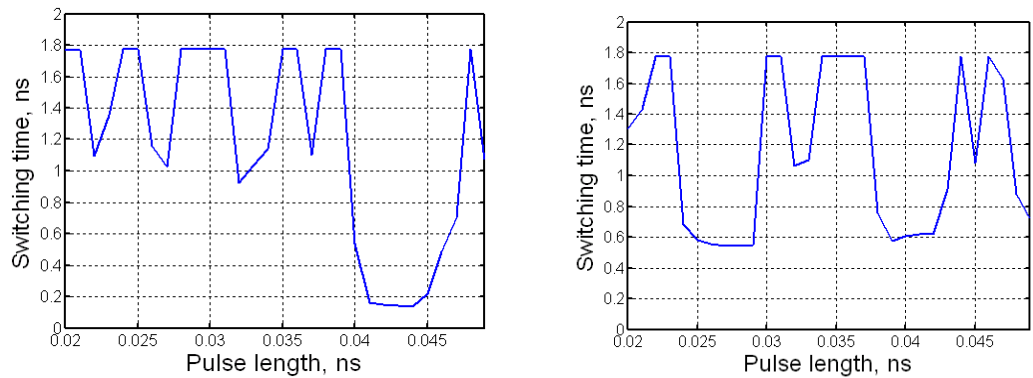

Figure 16. Same as Fig. 13 for two pulses with total duration of 0.2ns, at I=4mA, field-like torque 0.3, α=0.01, and polarization angle perpendicular to plane of 90 degrees. Simulation time 1.8ns.